\documentclass[12pt, a4paper]{article}

\pdfoutput=1

\usepackage{amsmath}
\usepackage{array}
\usepackage{amsfonts}
\usepackage{amssymb}
\usepackage{graphicx, rotating}
\usepackage{epsfig}
\usepackage{latexsym}
\usepackage{graphicx}
\usepackage{color}
\usepackage{amsmath,bm,amssymb}
\usepackage{cite}
\usepackage{slashed}
\usepackage{hyperref}

\newcommand{\lp}{\left(}
\newcommand{\rp}{\right)}
\newcommand{\nn}{\nonumber}
\newcommand{\be}{\begin{equation}}
\newcommand{\ee}{\end{equation}}
\newcommand{\bea}{\begin{eqnarray}}
\newcommand{\eea}{\end{eqnarray}}

\begin{document}

\begin{titlepage}
\begin{flushright}
DESY 18-162\\
\end{flushright}
\vspace{.3in}

\vspace{1cm}
\begin{center}
{\Large\bf\color{black}
Bubble wall velocities in the Standard Model \\ and beyond
}\\
\bigskip\color{black}
\vspace{1cm}{
{\large G.~C.~Dorsch$^1$, S.~J.~Huber$^2$, T.~Konstandin$^1$}
\vspace{0.3cm}
} \\[7mm]
{\it 
{$^1$ DESY, Notkestra\ss Ÿe 85, D-22607 Hamburg, Germany} \\
{$^2$ Department of Physics and Astronomy, University of Sussex, Brighton BN1 9QH, UK} \\
}
\end{center}
\bigskip

\vspace{.4cm}

\begin{abstract}
We present results for the bubble wall velocity and bubble wall thickness during a cosmological first-order phase transition in a condensed form. 
Our results are for minimal extensions of the Standard Model but in principle are applicable to a much broader class 
of settings. Our first assumption about the model is that only the electroweak Higgs is obtaining a vacuum expectation value during the phase transition. The second is that most of the friction is 
produced by electroweak gauge bosons and top quarks. Under these assumptions the bubble wall velocity and thickness can be deduced as a function of two equilibrium properties of the plasma: the strength of the phase transition and the 
pressure difference along the bubble wall.
\end{abstract}

\bigskip

\end{titlepage}

\section{Introduction \label{sec:intro}} 

The main property of a first-order phase transition is that it proceeds by bubble nuclation: Small regions of the new phase expand into a sea of old phase. In cosmology, such a phase transition can lead to many interesting phenomena, for example baryogenesis~\cite{
Kuzmin:1985mm, Morrissey:2012db, Konstandin:2013caa}, 
gravitational wave production~\cite{
Witten:1984rs, Kosowsky:1991ua, Kosowsky:1992rz, Kosowsky:1992vn, Kamionkowski:1993fg, Caprini:2015zlo} 
or generation of magnetic fields~\cite{
Vachaspati:1991nm}. 
One important characteristic is hereby the velocity with which the bubble interface moves. The bubble wall velocity has to be subsonic for efficient baryogenesis, while larger wall velocities typically lead to an enhanced production of gravitational waves. 

Some characteristics of the phase transition only rely on the equilibrium properties of the plasma, for example the latent heat or nucleation probability of critical bubbles. Other characteristics depend on the hydrodynamic behavior, for example the energy budget of the phase transition (much latent heat is transferred into bulk motion of the plasma versus heating the plasma).
The wall velocity falls into a third category that even depends on the microscopic properties of the a plasma. Friction is generated by the particle species that change mass during the phase transition and are driven out-of-equilibrium by the bubble interface. 
Quantifying the friction in the wall and bubble wall velocity hence requires a knowledge not only about the scalar sector of the theory (that releases the latent heat) but also about the particles that cause the friction~\footnote{In principle, a phenomenological model can be used to bypass this~\cite{Huber:2013kj, Leitao:2014pda}.}. Because of this, friction in the Standard Model~\cite{Moore:1995ua, Moore:1995si} works for example quite differently than in extensions of the Standard Model, e.g.~in the MSSM~\cite{John:2000zq} or in models with extended scalar sectors~\cite{Kozaczuk:2015owa}.

The present work aims at presenting results for the bubble wall velocity and the bubble wall thickness for the electroweak phase transition in a large class of models. The main assumption that is used to achieve this is the following: Motivated by the absence of BSM at collider experiments so far, we assume that only the $W$-bosons and top quarks dominate the friction. Any particle that contributes to friction requires a strong coupling to the Higgs. In turn, a strong coupling to the Higgs can leave traces in the production and decay rates of the Higgs, which is at odds with measurements at LHC. So, actually this assumptions seems well motivated. 
Even under this assumption, it is not easy to present the results on wall velocity since it depends 
in principle on three parameters. In order to make progress, we impose the condition that the nucleation probability is the correct one for a phase transition at electroweak scales. This allows us to remove one parameter and to produce results depend only on two parameters and are easy to digest. 

The structure of the paper is as follows: In Section~\ref{sec:fric} we outline the basic calculation of friction in the bubble wall. Section~\ref{sec:tunneling} discusses how to reduce the number of input parameters by using the tunneling probability as a universal (model-independent) constraint.
In Section~\ref{sec:run} we discuss relativistic bubble wall velocities before we 
present our main results in Sec.~\ref{sec:res}. Finally, we comment on phase transition with 
several scalar fields in Sec.~\ref{sec:multi} and conclude in Sec.~\ref{sec:diss}.

\section{Friction \label{sec:fric}} 

We will calculate the wall friction following the approach of Refs.~\cite{Moore:1995ua, Moore:1995si} and \cite{Konstandin:2014zta}. We will be rather brief here. More details on the calculation can be found in these two references. The equation of motion for the particles in the plasma is the Boltzmann equation
\be
p^\mu \partial_\mu \, f_i  = {\rm collisions + forces} \, .
\ee
The forces in the plasma are produced by the fact that the particles change their mass 
through their coupling to the Higgs VEV that changes in the bubble wall.
In the wall frame, the dispersion relation, $p^2 = m(z)^2$, is respected by the (pseudo-) particles in the plasma by 
changing the $p_z$ momentum. This results (see eq.~(5) in~\cite{Konstandin:2014zta}) in the forces 
\be
{\rm forces} = -\frac12 \partial_z m_i^2 \partial_{p_z} f_i \, ,
\ee
in the Boltzmann equation.

Typically, two-by-two scattering processes are so fast in the plasma that the system attains kinetic equilibrium on scales smaller than the wall thickness. In this case, it is justified to use for the particles in the plasma the flow {\em Ansatz}
\bea
f_i &=& \frac{1}{\exp(X_i) \pm 1} \, , \nn \\
X_i &=& \beta_i [ u^i_\mu p^\mu + \mu_i ] \, ,
\eea
where $u^i_\mu = \gamma_w (1,0,0,v_w)$ is the (local) plasma four-velocity, $\beta_i$ denotes the inverse temperature and $\mu_i$
the chemical potential. Particle number changing interactions are usually slower than two-by-two scattering processes such that the system will eventually relax to equilibrium ($\mu = 0$) away from the bubble wall. 
Notice that the forces inject energy and momentum into the fluid such that the four-velocity and the temperature do not coincide on both sides of the wall. Moreover, in the case of subsonic wall velocity, a shock builds up in front of the wall such that the temperature in front of the wall does not coincide with the nucleation temperature. We will take this effect also into account in our analysis.

As long as the system is relatively close to equilibrium, the deviations from equilibrium (in the symmetric phase)
can be parametrized by small deviations 
\be
X = \beta [ u_\mu p^\mu + \mu ] \equiv (u^\mu + \delta u^\mu(z) + \delta \tau(z)) \beta \, p_\mu + \delta \mu(z) \, .
\ee
Here, we defined the dimensionless deviations $\delta \tau = -\delta T/T$ and $\delta \mu = \mu/T$.
Normalization of the four-velocity ($u^\mu \delta u_\mu = 0$) then leads to three degrees of freedom per particle species in the plasma. The most important species in the Standard Model are hereby the top quark and the $W$-boson.
As long as there are no additional light degrees of freedom that couple strongly to the Higgs field, friction 
will be dominated by these two species. 

Using this {\em Ansatz} the Boltzmann hierarchy can be truncated and it remains to solve three equations 
from the hierarchy to determine the three deviations $\{\delta \mu, \delta \tau, \delta u^\mu\}$.
A natural choice is to use current and energy-momentum conservation 
of the different species in the plasma
\bea
\label{eq:plasm_eom}
\partial_\mu J_i^\mu &=& {\rm collisions} \, , \nn \\
\partial_\mu T_i^{\mu\nu} &=& {\rm collisions + forces} \, . 
\eea
In the wall frame the energy-momentum tensor and current are given in terms of the particle distribution functions as 
\bea
J_i^{\mu}(z) &=& \left. \int \frac{d^3p}{(2\pi)^3} \frac{p^\mu}{p_0} \, 
f_i(\vec p , z) \right|_{p_0 = \sqrt{\vec p^2  + m_i^2}} \, , \nn \\
T_i^{\mu\nu}(z) &=& \left. \int \frac{d^3p}{(2\pi)^3} \frac{p^\mu p^\nu}{p_0} \, 
f_i(\vec p , z) \right|_{p_0 = \sqrt{\vec p^2  + m_i^2}}  \, .
\eea
The forces arise hereby from the change in dispersion relation of the particles in the plasma as discussed above
\be
{\rm forces} = \frac12 \partial_\nu m^2(z) \, 
\int \frac{d^3p}{(2\pi)^3} \frac{1}{E} f(\vec p, z) \, .
\ee
Note that the equation of the current does not contain a force since the corresponding integral by construction vanishes~\cite{Konstandin:2014zta}. This is consistent with the picture that the force corresponds to a kinematic effect that changes the momentum of the particles but does not change particle numbers.
Note also that by construction the deviations from equilibrium vanish in the wall frame in case of a static wall, {\em i.e.}~when $\partial_\mu \to \partial_z$ and $u_z=0$. This is due to a cancellation between the force term and 
the $z$-dependence in the energy-momentum tensor through the mass. One way to see this 
is to observe that the Boltzmann equation in the wall frame can be written in terms of the Poisson bracket 
\be
\left\{ a , b \right\} = \frac{da}{dz} \frac{db}{dp_z} - \frac{da}{dp_z} \frac{db}{dz} \, ,
\ee
and then reads 
\be
\label{eq:BasP}
\frac12 \left\{ \vec p^2 + m^2 , f \right\} = {\rm collisions} \, .
\ee
For a static wall, the equilibrium distribution only depends on $E^2 = \vec p^2 + m^2$ and hence solves the Boltzmann
equation. 

Ultimately, the equations are linearized in the deviations from equilibrium and are of the form~\cite{Konstandin:2014zta} 
\be
\label{eq:linarized}
A \cdot q^\prime + \Gamma \cdot q = S \, ,
\ee
where $A$ contains velocity dependent moments of the equilibrium distributions, $\Gamma$ contains scattering rates from particle number changing interactions, $q$ contains the deviations from equilibrium $\{\delta \mu, \delta \tau, \delta u^\mu\}$ for all relevant species and the background of light particles and $S$ is the source term from the forces. Notice also that the prime denotes $\gamma_v \partial_z$ in the wall frame with the Lorentz factor of the wall velocity. This will become relevant in the runaway regime.

\vskip 0.3 cm

The second equation of motion on has to solve is for the Higgs field. This can be obtained by energy-momentum 
conservation of the full system Higgs+fluid. This equation reads
\be
\Box \phi + \frac{dV}{d\phi} + \sum_i \frac{dm^2_i}{d\phi}
\int \frac{d^3p}{(2\pi)^3} \frac{1}{2E} f_i(\vec p, z) = 0 \, . 
\ee
The last term is again the force term resulting from the impact of the dispersion relation on the 
particle kinematics. An alternative way of writing this equation is by splitting the fluid part into an equilibrium piece and 
a deviation from equilibrium. The equilibrium piece is the free energy of the fluid and
the equation hence reads 
\be
\label{eq:phi_eom}
\Box \phi + \frac{dV_T}{d\phi} + \sum_i \frac{dm^2_i}{d\phi}
\int \frac{d^3p}{(2\pi)^3} \frac{1}{2E} \delta f_i(\vec p, z)   = 0 \,, 
\ee
where $V_T$ denotes the thermal effective potential (or more precisely the free energy of the system).
The last contribution involving the deviation from equilibrium $\delta f$ can then be identified as the friction produced by the wall. Notice that this construction
is not unique since one can choose as equilibrium either the phase in front of the wall or the phase behind the
wall that are at different temperature. 

Again, to solve this equation we use an {\em Ansatz} for the Higgs VEV profile, namely a $\tanh$ profile with 
a wall thickness $L_w$. The final Higgs equations then read
\bea
\label{eq:eom_Higgs}
\frac{\Delta V_T}{T^4} &=& f \, , \nn \\
-\frac{2}{15(TL_w)^2} \lp \frac{\phi_0}{T}\rp^3 + \frac{W}{T^5} &=& g \, ,
\eea
where $\Delta V_T$ denotes the difference in effective potential (free energy) between the two phases, $\phi_0$ denotes the 
Higgs VEV in the broken phase~\footnote{
Notice that these quantities are evaluated right in front of the wall. For deflagrations, there is a difference to the plasma in the symmetric phase due to the shock front before the bubble wall
interface that we take into account.}
, $T = 1/\beta$ is the temperature and $W$ denotes the following integral 
\be
W \equiv - \int_0^{\phi_0} \frac{dV_T}{d\phi} (2\phi - \phi_0) d\phi \, .
\ee
While $\Delta V$ is the pressure along the wall, $W$ is a measure for the pressure gradient along the wall
and mostly determines the wall thickness. 
The functions $f$ and $g$ are the corresponding integrals over the force term in the Higgs equation of motion
\bea
\label{eq:DV}
f &=&  T^{-4}\int dz \frac{d\phi}{dz} \sum_i \frac{dm^2_i}{d\phi} 
\int \frac{d^3p}{(2\pi)^3} \frac{1}{2E} \delta f_i(\vec p, z) \, , \\
\label{eq:W}
g &=&  T^{-5}\int dz \frac{d\phi}{dz} (2\phi - \phi_0) \sum_i \frac{dm^2_i}{d\phi} 
\int \frac{d^3p}{(2\pi)^3} \frac{1}{2E} \delta f_i(\vec p, z) \, . 
\eea
The two dimensionless functions $f$ and $g$ have to be obtained by solving the fluid system~(\ref{eq:plasm_eom}).
They depend on the strength of the phase transition $\phi_0/T$, the wall velocity $v_w$, the wall thickness $L_w$ and several couplings and collision rates in the top quark and $W$-boson  sectors~\cite{Konstandin:2014zta}. 
Aim of the present paper is to disentangle this dependence and provide an easy to use result for the bubble wall velocity that applies to a wide range of models.

\vskip 0.3 cm

In the present approach we have used the assumptions that the system is locally in kinetic equilibrium and that the deviations from equilibrium are small enough to allow for the flow {\em Ansatz}. Clearly, this is not always justified. One important example are wall velocities close to the speed of sound. This can lead to large deviations from equilibrium even when the forces are small. This is signaled by a singular matrix $A$ in (\ref{eq:linarized}) when the system in linearized in the deviations $\{\delta \mu, \delta \tau, \delta u^\mu\}$. Another notable regime are relativistic bubble wall velocities. In this case scattering processes are suppressed and the system does not even attain kinetic equilibrium in the wall.
This bound is discussed in detail in section~\ref{sec:run}.

\vskip 0.3 cm

A common strategy to solve these equations is to solve the fluid system for fixed wall velocity $v_w$ and wall thickness $L_w$ and then to vary these two parameters until the Higgs equations (\ref{eq:DV}) and (\ref{eq:W}) are fulfilled. In most cases, the function $g$ plays a subdominant role and the wall thickness is determined by $W$ only (for fixed wall velocity). At the same time, the function $f$ is proportional to the wall velocity what ultimately determines the 
expansion speed of the Higgs bubbles. Finally, notice that in case of subsonic walls, there is a shock in front of the wall such that the temperature in front of the wall does not coincide with the nucleation temperature of the
system. This mismatch depends on the wall velocity~\cite{Espinosa:2010hh} and we take this effect into account.

\section{Tunneling as a constraint \label{sec:tunneling}} 

As discussed in the last section, to determine the wall velocity (or its friction), one has to solve the dynamics of
the fluid in conjuncture with the Higgs equation of motion. For a concrete model, the nucleation temperature and 
corresponding effective potential are known. This means that the strength of the phase transition,
$\phi_0/T$, as well as the quantities $\Delta V$ and $W$ in (\ref{eq:eom_Higgs}) are specified. So it remain the wall velocity $v_w$ and the wall thickness $L_w$ that have to be fixed by fulfilling (\ref{eq:eom_Higgs}) using the deviations from equilibrium $f$ and $g$ obtained from solving (\ref{eq:linarized})~\footnote{As mentioned earlier, for subsonic walls a shock precedes the bubble which adds one more complication in the procedure. However, the shock only depends on the wall velocity and hence this part does not depend on the model and can be easily incorporated. }.

For every set of concrete model and input data $\{\Delta V, W, T, \phi_0\}$ these equations can be solved. 
Ultimately, the problem is that the general solution cannot be visualized easily to make it applicable for general models. Also the functions $f$ and $g$ (even in the simplest case of a SM particle content) still depend on three quantities: The wall velocity $v_w$, the wall thickness $L_w$ as well as the strength of the phase transition $\phi_0/T$. Hence, also these functions cannot be easily visualized and/or parametrized.

Our main idea is to reduce the number of relevant parameters from four to two. More concretely
\be
\{\Delta V, W, T, \phi_0\} \to \{\Delta V/\phi_0^4, \phi_0/T \} \, .
\ee
This makes it possible to provide contour plots for the wall velocity $v_w$ and wall thickness $L_w/T$
that are valid for a wide range of models. First, since we are only interested in the dimensionless 
quantities $v_w$ and $L_w/T$, trivially only three dimensionless combinations of the four parameters can enter. To reduce the number of parameters further to two requires a bit more work.

In order to achieve this, we use the tunneling action to constrain $W$ in terms of $\Delta V$. To get the three-dimensional tunnel action~\cite{Coleman:1977py,Callan:1977pt,Linde:1980tt}, one has to calculate the bounce solution that obeys 
\be
\frac{d^2\phi}{d\rho^2} + \frac{2}{\rho}\frac{d\phi}{d\rho} = \frac{dV}{d\phi} \, .
\ee
This equation has the following properties: If the potential is scaled, $V \to \lambda V$, the bounce solution scales as $\rho \to \lambda^{-1/2} \rho$ and accordingly $S_3 \to \lambda^{-1/2} S_3 $. Likewise, if the potential is stretched, $\phi \to \lambda \phi$, the bounce solution scales as $\rho \to \lambda \rho$ and accordingly $S_3 \to \lambda^3 S_3 $. So, the tunneling action for any family of potentials with three free parameters can be written as 
\be
\label{eq:tunneling}
\frac{S_3}{\phi_0} = \frac{W^{3/2} \phi^{1/2} }{\Delta V^2} \times X( W / \Delta V / \phi_0)\, ,
\ee
where the argument $W / \Delta V / \phi_0$ is invariant under above rescalings. 
The prefactor is chosen such that it reproduces the correct behavior of $S_3 \to \lambda^{-1/2} S_3 $ and such that in the thin wall regime $W / \Delta V / \phi_0 \to \infty$ the function $X( W / \Delta V / \phi_0)$ approaches a constant.

\begin{figure}[h!]
\begin{center}
\includegraphics[width=0.7\textwidth]{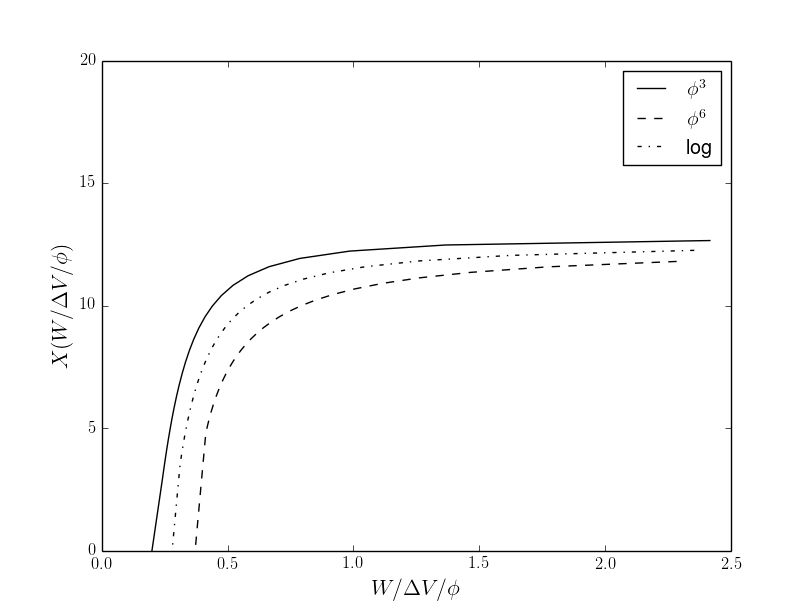}
\end{center}
\caption{\label{fig:S3ophi_models}
The plot shows the function $X$ defined in (\ref{eq:tunneling}). The limiting cases are the models with a $\phi^3$ and $\phi^6$ terms in the free energy that should cover most of realistic potentials. }
\end{figure}

The question is how different 
$X( W / \Delta V / \phi_0)$ can be across different families of potentials. Figure~\ref{fig:S3ophi_models} shows this function for three families  of potentials. They all share the operators $\phi^2$ and $\phi^4$. In order to 
construct a barrier, a third operator is necessary and this is the cubic ($\phi^3$), dimension six ($\phi^6$) or logarithmic ($\phi^4 \log(\phi)$) operator. Notice that there is no reference to temperature in this figure. The potential are not realistic finite temperature potentials but the three examples should exemplify the range of possibilities that arise at nucleation temperature. 

For most parts, the models agree for 
\be
\label{eq:WVbound}
W > 0.5 \, \Delta V \, \phi_0 \, . 
\ee
This is the most relevant regime for several reasons. Observe that this is the thin wall regime where the barrier is relatively small compared to the 
potential difference. 
Furthermore, (\ref{eq:tunneling}) can be written as 
\be
\frac{\sqrt{\Delta V}}{\phi^2} = \frac{\xi X}{(S_3/T)} \lp \frac{W }{\Delta V \phi} \rp^2 \, .
\ee
Since $S_3/T \simeq 140$ and $X<15$, a small $W/\Delta V \phi$ is in most models in tension with the observed Higgs mass.  We will display the bound (\ref{eq:WVbound}) in our final results.

The relation (\ref{eq:tunneling}) relates $\Delta V$ and $W$ and the wall velocity $v_w$ and the wall thickness $L_w$ can be expressed as a function of $\Delta V/\phi_0^4$ and $\phi_0/T$ only. As discussed above, this should work 
especially well in the limit of small wall velocities and thick walls. Hence, we will use the opposite regime to benchmark the quality of our results, in particular models with runaway walls. 

\section{Runaway \label{sec:run}} 

The regime of relativistic wall velocities was first discussed in Ref.~\cite{Bodeker:2009qy} and a higher order effect was presented in Ref.~\cite{Bodeker:2017cim}.
In the regime of relativistic wall velocity the system becomes collisionless [remember the Lorentz factor in (\ref{eq:linarized})]. More quantitatively, this happens for 
\be
\label{eq:fluidConstraint}
\Gamma \, L_w \ll \gamma_w \, A \, ,
\ee
where $A$ contains velocity dependent moments of the equilibrium distributions (that are of order one for large velocities), $\Gamma$ contains scattering rates from particle number changing interactions, $L_w$ is the wall thickness and $\gamma_w$ is the Lorentz factor coming from the wall velocity.

In this regime, the $z$-momentum changes according to the dispersion relation. Since the system is 
collisionless, this allows to directly infer the particle distribution function behind the wall
in the broken phase from the one in the symmetric phase
\be
\label{eq:shift}
f_b(p_z) = f_s(\bar p_z) \, ,\quad  \bar p_z = \sqrt{p_z^2 -m_b^2 +m_s^2} \, .
\ee
Notice that this reasoning is consistent wit the collisionless Boltzmann equation in (\ref{eq:BasP}).
Moreover, the energy momentum tensor in the broken phase can be evaluated by undoing the shift in 
$p_z$ in the integration variable. In particular one finds
\bea
J^{z}_b &=& J_s^{z}\, , \nn \\
T^{0z}_b &=& T_s^{0z}\, , \nn \\
T^{zz}_b &=& T_s^{zz} + \Delta T_{BM}^{zz} \, ,
\eea
with
\be
\label{eq:meanfield}
\Delta T_{BM}^{zz} = (m_b^2 - m_s^2) \left. \int \frac{d^3p}{(2\pi)^3} \frac{1}{2E} f(E) 
\right|_{E = \sqrt{\vec p^2 + m_s^2}} \, .
\ee
More generally, any moment that contains one factor $p_z$ and arbitrary factors of $E$ will be conserved in the wall, just like $J^z$ and $T^{0z}$. This is due to the fact that $dp_z \, p_z$ does not change under the 
shift in (\ref{eq:shift}).

In case the particle is massless in the symmetric phase, the pressure in the runaway regime is the mean-field contribution (\ref{eq:meanfield}) to the effective potential using the particle distribution in the symmetric phase.
While the Boltzmann equation respects this behavior, the fluid {\em Ansatz} does not. In particular, after linearization, the energy-momentum tensor fulfills 
\be
\label{eq:equilibr}
\partial_z T^{zz} = \partial_z m^2(z) \left. \int \frac{d^3p}{(2\pi)^3} \frac{1}{2E} f(E) 
\right|_{E = \sqrt{\vec p^2 + m(z)^2}} \, ,
\ee
which underestimates the true pressure difference. Hence we expect that the fluid {\em Ansatz} overestimates the
wall velocity in this regime. This difference can be quite substantial. In the case of the top and $W$-bosons, the particle masses can be significantly larger than the temperature such that (\ref{eq:meanfield}) and (\ref{eq:equilibr}) differ by a factor of order unity.

In terms of $\Delta V$, the correct condition for runaway is then
\be
\label{eq:runaway_bound}
\frac{\Delta V}{\phi_0^4} > \frac{\Delta T_{BM}^{zz} - V_T}{\phi_0^4}\, . 
\ee
In our setup, only the top quark and $W$-boson are taken into account and this bound becomes independent from
the actual zero temperature scalar potential because the right-hand-side of (\ref{eq:runaway_bound})
can be calculated as a function of $\xi = \phi_0/T$ and the involved couplings. We will present the runaway bound (\ref{eq:runaway_bound}) and the constraint from the breakdown of the fluid approximation (\ref{eq:fluidConstraint}) along the results for the 
wall velocity in the next section.

\section{Results \label{sec:res}} 

\begin{figure}[h!]
\begin{center}
\includegraphics[width=0.7\textwidth]{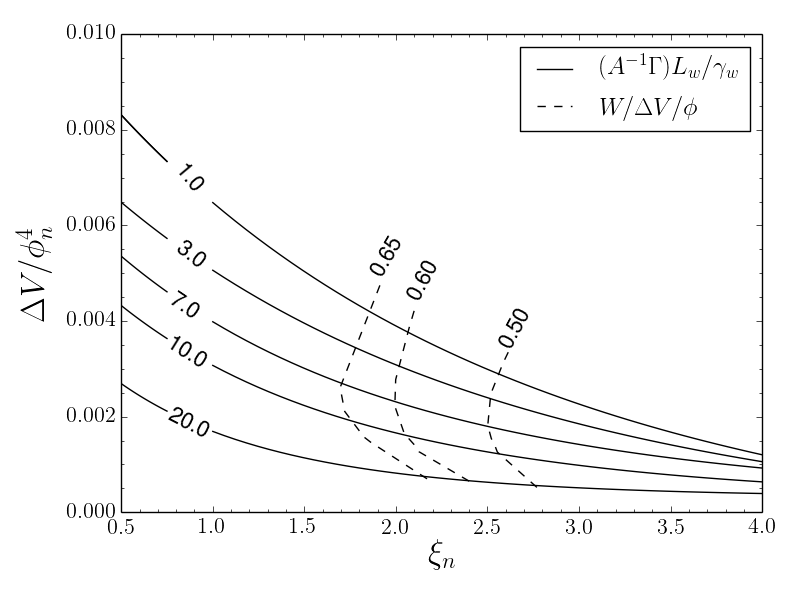}
\includegraphics[width=0.7\textwidth]{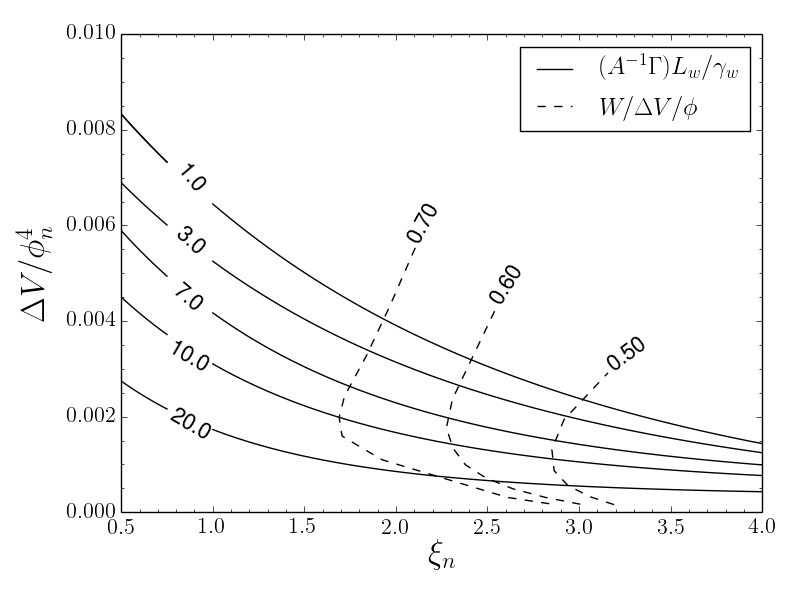}
\end{center}
\caption{\label{fig:constraints} The constraints (\ref{eq:WVbound}) and (\ref{eq:fluidConstraint}) as a function of the phase transition strength $\xi_n$ and the normalised vacuum energy $\Delta V/\phi_n^4$, for the (top) cubic toy-model and (bottom) dimension six extension.}
\end{figure}

For the computation of the wall velocities from the fluid equations~(\ref{eq:eom_Higgs}) we consider two different potentials, namely a SM extension with a $\phi^6$ operator~\cite{Bodeker:2004ws} and a toy model with a tree-level cubic term. The tree-level potentials are, respectively,
\begin{equation}
	V_{\rm tree}^{\phi^6} = -\frac{\mu^2}{2} \phi^2 + \frac{\lambda}{4} \phi^4 + \frac{\phi^6}{8\,M^2}\, ,
\end{equation}
\begin{equation}
	V_{\rm tree}^{\phi^3} = \lambda\,\phi^2 (\phi-\phi_0)^2 - D\,T_0^2\,\phi^2 \, ,
\end{equation}
with $D \equiv \frac{1}{24\,v^2} \left( 6 m_t^2 + 6 m_W^2 + 3 m_Z^2 \right)$ the coefficient of the quadratic term in the thermal potential for the SM particle content. In the $\phi^6$ case we enforce that the minimum be at $v=246.22$~GeV, varying the scalar mass and the overall cutoff scale $M$. For the $\phi^3$ we fix the overall scale $T_0=100$~GeV and vary $\phi_0$ and $\lambda$. On top of the tree-level potentials, we also add the one-loop Coleman-Weinberg as well as the thermal contributions.

\subsection{Constraints from the universality of tunneling}

As seen in Fig.~\ref{fig:S3ophi_models} the behaviour of the tunneling action is quite universal in terms of $W/\Delta V/\phi_0$ as long as the phase transition is relatively weak. However, in the regime of strong phase transitions, the models start to differ. In particular, the barrier in the effective potential vanishes at 
different values of $W/\Delta V/\phi_0$ for different models (sending $S_3/T$ to zero). 
Results for $W/\Delta V/\phi_0$ for above two models are shown in Fig.~\ref{fig:constraints}
We impose 
\be
\label{eq:Wbound}
W/\Delta V/\phi_0 > 0.5 \, ,
\ee
in our final results. 

\subsection{Constraints from the fluid approximation}

Another approximation that can break down is the fluid approximation. The fluid approximation requires that the interactions are strong enough to keep the system in kinematic equilibrium. In our fluid system this implies
\be
\label{eq:Abound}
(A^{-1}\Gamma) \,  L_w > \gamma_w \, .
\ee
Results for this bound are also displayed in Fig.~\ref{fig:constraints}.
If this relation is violated, the friction calculated will not be accurate. We will see this explicitly in the 
runaway regime that occurs in a regime of parameter space that violates this bound.

\subsection{Final results}

\begin{figure}[h!]
\begin{center}
\includegraphics[width=0.7\textwidth]{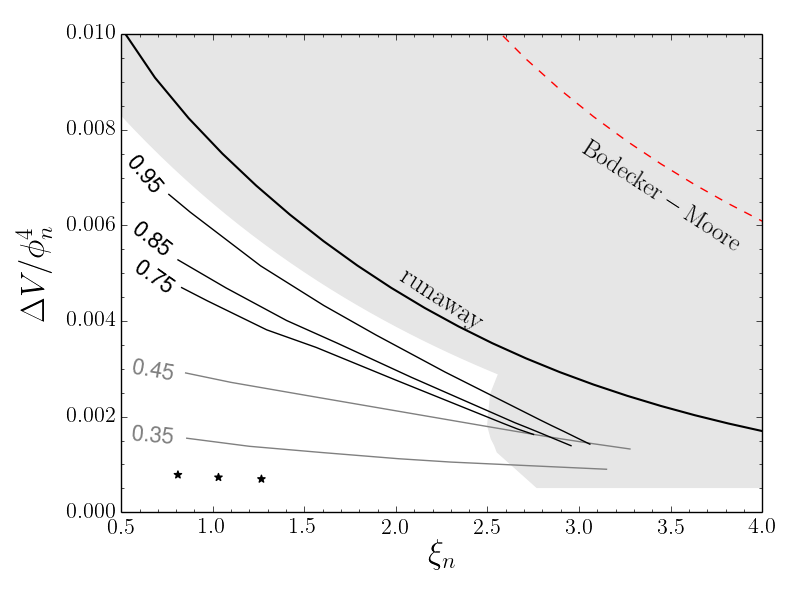}
\includegraphics[width=0.7\textwidth]{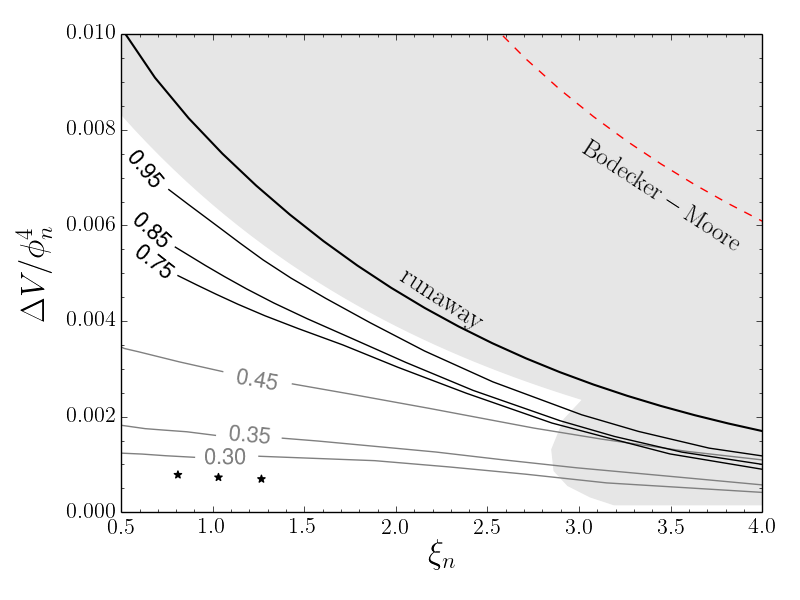}
\end{center}
\caption{\label{fig:moneyplots} The plot shows the wall velocity as a function of the phase transition strength $\xi_n$ and the normalised vacuum energy $\Delta V/\phi_n^4$, for the (top) cubic toy-model and (bottom) dimension six extension. The stars denote the Standard Model with a very light Higgs boson with masses $m_H \in \{50,30,20\}$ GeV. The results in the shaded regions are unreliable as seen in Fig.~\ref{fig:constraints}.}
\end{figure}

The results are shown in Fig.~\ref{fig:moneyplots}. We plot the wall velocity as a function of the phase transition strength $\xi_n$ and the normalised vacuum energy $\Delta V/\phi_n^4$, both evaluated at the nucleation temperature. Also shown is the curve obtained from application of the B\"odecker-Moore criterion. 
The shaded region results from violation of one of our requirements in (\ref{eq:Wbound}) and (\ref{eq:Abound}).

Note that the curves in the two plots essentially coincide in the region of mildly strong phase transitions, $\xi_n \lesssim 2.7$, in agreement with the discussion of Fig.~\ref{fig:S3ophi_models}. For stronger transitions we approach the case $W/\Delta V/\phi \lesssim 0.5$, where the nucleation criterion alone does not allow us to draw model-independent conclusions based only on $\Delta V$ and $\xi_n$.

Another important result presented in the plot is the huge gap between the model-independent runaway curves obtained from the B\"odecker-Moore criterion and from the flow \emph{Ansatz}. As the shaded regions indicate, the discrepancy seems to stem from a breakdown of the fluid approximation, as collisions become too large for the perturbations to be treated as small. Notice that also this disparity shows up in the shaded region due to the lack of sufficient interactions to stay close enough to equilibrium.

Notice that our results for the Standard Model are quite off from what was originally found in~\cite{Moore:1995si}. As already noted in~\cite{Konstandin:2014zta} this is due to the fact that 
the infrared divergence in the bosonic force terms is aleviated by using the full mass dependence (and to a much lesser extent that we take into account the shock front before the wall in a non-linear way - an effect that is only relevant for strong phase transitions). 

\begin{figure}[h!]
\begin{center}
\includegraphics[width=0.7\textwidth]{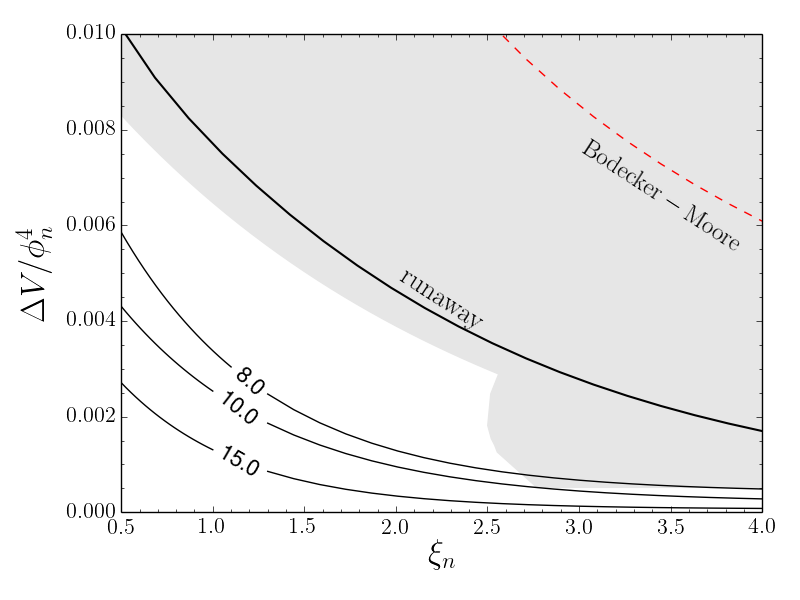}
\includegraphics[width=0.7\textwidth]{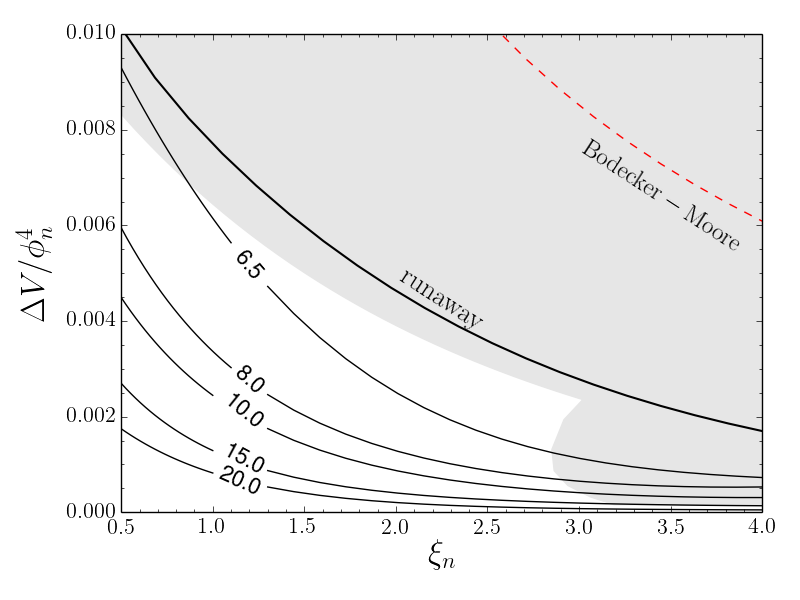}
\end{center}
\caption{\label{fig:moneyplots2} The same as Fig.~\ref{fig:moneyplots} for the wall thickness.}
\end{figure}

Finally, notice that subsonic and supersonic solutions cross in the cubic model. This is due to the shock front that leads to an ambiguity in $\Delta V/\phi^4$ evaluated at the nucleation temperature. If it was evaluated at the bubble wall, this ambiguity would be absent.

\section{Phase transitions with several scalars \label{sec:multi}} 

Before concluding, we comment on the case where several scalar fields are involved in the phase transition. As a prototype of this situation we have in mind the Standard Model enhanced by a singlet that can develop a two-stage phase transition~\cite{Espinosa:2011ax}. The second stage will break the electroweak symmetry and can be very strong due to a potential barrier that is even present at tree level. 

The Higgs field $\phi$ and the singlet field $s$ will both vary during the phase transition.
The fields will hence describe a path $\{\phi(z),s(z)\}$ in scalar field space. This path will pass through the two minima and surpass the barrier close to a saddle point of the potential.
In case the path was known, one could reduce the problem to a one-dimensional problem. Consider a variable $p$ that parametrizes the path traveled 
\be
d\phi^2 + ds^2 = dp^2 \, .
\ee
The equation of motion for $p$ is obtained by summing over the corresponding equations for $\phi$ and $s$ of the form (\ref{eq:phi_eom}). Multiplying these equations by $d\phi/dz$ and $ds/dz$, respectively and summing over the integration one finds again an equation of the form 
\bea
\label{eq:eom_Higgs_multi}
\frac{\Delta V_T}{T^4} &=& f \, ,  \\
\label{eq:eom_Higgs_multi_W}
-\frac{2}{15(TL_w)^2} \lp \frac{p_0}{T}\rp^3 + \frac{W}{T^5} &=& g \, ,
\eea
with the contour integral
\be
W \equiv - \int_0^{p_0} 
\lp \frac{dV_T}{d\phi} d\phi + \frac{dV_T}{ds} ds\rp
 (2p - p_0)  \, .
\ee
The expression for $f$ is unchanged while the definition $g$ will contain a factor $(2p-p_0)$ instead of $(2\phi-\phi_0)$. In any case, the function $g$ is less relevant and typically smaller than the other two contributions to the second equation of (\ref{eq:eom_Higgs_multi}). 

We would like to understand to what extent it is possible to reuse the results of the one scalar case for potentials with several scalars. Unfortunately, this is very limited. Since $f$ and $g$ are rather involved functions of $v_w$ and $L_w$ and $\phi/T$, the only way to make this happen is to map $\phi_0/T$ in the one scalar case to $\phi_0/T$ in the multi-scalar case and check if this results in a simple transformation on $\Delta V$. According to (\ref{eq:eom_Higgs_multi}) also $\Delta V$ has to be unchanged.

Since $p$ is canonically normalized, the tunneling analysis is basically unchanged and we can still express $\Delta V$ as a function of $W$. For fixed $\Delta V$ and $S_3/T$, one finds the scaling $W \propto p^{-1}$ according to (\ref{eq:tunneling}). This means in turn, that the $W$ inferred in the single scalar case is a factor $p_0/\phi_0$ too large. At the same time, the first term in (\ref{eq:eom_Higgs_multi_W}) is a factor $(p_0/\phi_0)^3$ too large.

In the end, the multi-scalar case will lead to thicker walls and also higher wall velocities than just feeding the values for $\xi = \phi_0/T$ and $\Delta V/\phi_0^4$ into the single scalar equations.

\section{Discussion \label{sec:diss}} 

We presented results for the bubble wall velocity and bubble wall thickness for the electroweak phase transition in case it is first order, see Figs.~\ref{fig:moneyplots} and \ref{fig:moneyplots2}. We discussed in detail under what assumptions these results can be used in extensions of the Standard Model. Using the requirement that the bubble nucleation probability is appropriate to trigger the phase transition, makes it possible to reduce the number of input parameters to two, for example the strength of the phase transition $\xi = \phi_0/T$ and the pressure difference along the wall $\Delta V/\phi_0^4$.

In order to quantify friction in the wall, we use the fluid approximation to model the out-of-equilibrium. Limitation of the method are either a lack of interaction to keep the plasma close enough to equilibrium. Besides, in order to reduce the number of input parameters, the bubble walls have to be sufficiently thin. Both limits are discussed in Fig.~\ref{fig:constraints}.

Finally, let us discuss to what extent our results might be extrapolated to more complicated situations. The main assumptions to our approach is that only the Higgs field is obtaining a VEV during the phase transition and that the friction is produced by a SM particle content, namely by the top quarks and electroweak gauge bosons. Extending our results to phase transitions where several scalar fields are relevant seems to be hindered by the fact that we cannot reduce the number of input parameters as before, see Sec.~\ref{sec:multi}. Similarly, changing the composition of the particles that are responsible for the friction is not easily accomplished. Even changing the particle content in a rather trivial way (e.g.~doubling the top and $W$-boson content or adding a very strongly/very weakly interacting species) will have an impact on the function $W$ that is non-trivial.

\section*{Acknowledgments}

This work of TK and GCD was supported by the German Science Foundation (DFG) under
the Collaborative Research Center (SFB) 676 Particles, Strings and the Early
Universe. The work of SJH is supported by the Science Technology and Facilities Council (STFC) under
grant  number  ST/P000819/1. 


\bibliography{wall_vel.bib}{}
\bibliographystyle{unsrt}

\end{document}